\documentclass[conference]{IEEEtran}
\IEEEoverridecommandlockouts
\usepackage{cite}
\usepackage{amsmath,amssymb,amsfonts}
\usepackage{algorithmic}
\usepackage{comment}
\usepackage{graphicx}
\usepackage{textcomp}
\usepackage{xcolor}
\def\BibTeX{{\rm B\kern-.05em{\sc i\kern-.025em b}\kern-.08em
    T\kern-.1667em\lower.7ex\hbox{E}\kern-.125emX}}
\newcommand{\system}{Serdab }
\begin{document}

\title{Serdab: An IoT Framework for Partitioning Neural Networks Computation across Multiple Enclaves 
}

\author{\IEEEauthorblockN{Tarek Elgamal}
\IEEEauthorblockA{\textit{University of Illinois Urbana-Champaign} \\
telgama2@illinois.edu}
\and
\IEEEauthorblockN{Klara Nahrstedt}
\IEEEauthorblockA{\textit{University of Illinois Urbana-Champaign} \\
klara@illinois.edu}
}

\maketitle

\begin{abstract}

Recent advances in Deep Neural Networks (DNN) and Edge Computing have made it possible to automatically analyze streams of videos from home/security cameras over hierarchical clusters that include edge devices, close to the video source, as well as remote cloud compute resources. However, preserving the privacy and confidentiality of users' sensitive data as it passes through different devices remains a concern to most users. Private user data is subject to attacks by malicious attackers or misuse by internal administrators who may use the data in activities that are not explicitly approved by the user. To address this challenge, we present Serdab, a distributed orchestration framework for deploying deep neural network computation across multiple secure enclaves (e.g., Intel SGX). Secure enclaves provide a guarantee on the privacy of the data/code deployed inside it. However, their limited hardware resources make them inefficient when solely running an entire deep neural network. To bridge this gap, Serdab presents a DNN partitioning strategy to distribute the layers of the neural network across multiple enclave devices or across an enclave device and other hardware accelerators. Our partitioning strategy achieves up to 4.7x speedup compared to executing the entire neural network in one enclave.





\end{abstract}


\section{Introduction}\label{sec:intro}

Internet of Things (IoT) applications generate massive amounts of real-time data. A large amount of this data comes from cameras. For example, estimates suggest the clusters of cameras on board of a smart vehicle are going to generate 3 TB of data per day and reports by Information Handling Services (IHS) indicate that 245 million professionally installed surveillance cameras are operating worldwide as of 2015~\cite{IHS}. Owners of such cameras strive to make predictions/inference from large streams of video feeds, often by deploying and serving deep neural network (DNN) models. However, due to the significant amount of data generated by IoT devices, it is infeasible to send the data to a centralized cloud and get the response in timely manner. For this reason, the concept of {\it edge computing}~\cite{cloudlet} has emerged. Edge computing refers to the computing and storage infrastructure that exists close to the sources of data.
An IoT application can be deployed on multiple hops of these edge devices where some analysis is done on a raspberry pi and the rest of the analysis runs in an edge data center. However, deploying IoT applications on the edge poses security and privacy challenges because users remain skeptical about the confidentiality of their input data and whether private user data is used by edge computing vendors in other applications.


To address the data confidentiality problem for edge/cloud computing, researchers proposed cryptographic approaches~\cite{cryptonets}\cite{oblivious} to enable privacy-preserving predictions. Although the performance of such approaches has significantly improved, they still remain impractical for production environments. On-device inference has also been proposed to protect data confidentiality~\cite{on-device1}\cite{on-device2} through processing parts of the application in the client's device and preventing the private data from leaving the client's device. However, such approaches consume a significant amount of computation and energy from the resource-constrained client devices. As an alternative, several systems~\cite{chiron}\cite{myelin} have proposed using trusted execution environments -also known as TEEs or enclaves- such as Intel Software Guard Extensions (SGX) to run machine learning workloads while preserving data confidentiality. In such systems, the private data is only decrypted within the trusted execution environment which is protected from all privileged software in the system such as operating system and virtual machine monitors. However, SGX-based computation is currently performance-and memory-constrained. For example, TEE workloads cannot exploit accelerated linear algebra libraries. It also has a limited memory size (128 MB) which limits the amount of computation that can be done within one TEE and it becomes essential to delegate part of the computation to run in another hardware accelerator (e.g., GPU, CPU, or another enclave) that sits in the same or a different edge device. Previous work has proposed different techniques to partition computation across enclave and a co-located hardware accelerator to improve the latency and privacy~\cite{slalom}\cite{yerba}. However, little work has been done on addressing the problem of partitioning computation across multiple enclaves that sit in geo-distributed edge devices. A key hurdle to adoption of this paradigm is the lack of a platform ecosystem that simplifies the deployment, and management of applications, seamlessly across multiple enclaves that sit in different devices. In this paper, we  present Serdab\footnote{An ancient Egyptian tomb structure in which secret chambers (enclaves) are connected with secret passages (communication channels)}, a distributed orchestration framework that allows deploying deep neural network computation across multiple enclaves.
A major challenge that \system addresses is how to optimally distribute deep neural network computation across an enclave and other devices (e.g., other enclaves or hardware accelerators). This is especially important considering the resource constraints of the enclave (e.g., 128 MB) that makes it inadequate to efficiently support running a deep neural network entirely within one enclave. An important question that we address in this paper is {\it how to partition the NN layers to reduce the overall latency of computation for a stream of video frames while maintaining data confidentiality}. To address this challenge we leverage an interesting insight that the output of intermediate layers of a convolutional neural network becomes dissimilar to the original input towards the last layers~\cite{vis}. For example, the output of layer 5 is  less similar to the original image compared to the output of layer 1. This insight can be used to run layers 1-5 inside the enclave to protect data confidentiality and offload the rest of the computation to an edge device that has a processor with more computing power (e.g., CPU or GPU). One disadvantage with this approach is that the majority of the layers might end up running in the enclave, leaving only a few layers to the faster processor. This problem is more critical in IoT applications because such applications typically require processing a stream of video frames and if most of the layers run on the enclave, the enclave will become the bottleneck and the entire application will be slowed down by the queuing time on the enclave, yielding the other processor idle waiting for the enclave. Our approach to solve this is to distribute the DNN layers across multiple enclaves. In this approach, the partitioning can happen at any layer because the intermediate data can only be decrypted in the next enclave. This helps the partitioning to be distributed equally across both enclaves which reduces the queuing delay, provides better resource utilization, and reduces the overall latency of processing a stream of frames. In this paper, we propose a method for partitioning Neural Network (NN) layers across multiple devices to process a stream of video frames. We compare our approach with different approaches for partitioning NN layers, and we discuss their applicability to IoT applications. In summary, we make the following contributions:

\begin{itemize}
    \item We propose a framework to compute IoT applications across multiple enclave devices or across an enclave device and other hardware accelerators.
    
    \item We propose a novel technique to find the best partitioning of neural networks computation across distributed resources. A key advantage in our neural network partitioning approach, that was not explored in previous work, is to leverage {\it pipeline parallelism} in which we consider that both enclaves concurrently execute different video frames.
    
    \item A complete end-to-end implementation and evaluation of the platform using a diverse set of videos that vary in the type of objects, resolution, and the geo-location in which the video was captured. Our results show that for a dataset of more than 10,000 video frames, our partitioning approach can achieve up to 4.7x speedup over the running the entire neural network in one enclave.

\end{itemize}

The rest of the paper is organized as follows. Section~\ref{sec:challenges} discusses background, motivation, and the threat model. In section~\ref{sec:system}, we describe our system architecture. We define the problem of privacy-aware placement in section~\ref{sec:models} and we present our technique to solve it in section~\ref{sec:approach}. We evaluate our system in section~\ref{sec:eval}. In section~\ref{sec:related}, we discuss the related work. Section~\ref{sec:conc} concludes the paper.

\section{Background and Motivation}\label{sec:challenges}
\subsection{Trusted Execution Environment (TEE)}

\begin{figure}[t]
      \centering
      \includegraphics[width=0.8\linewidth]{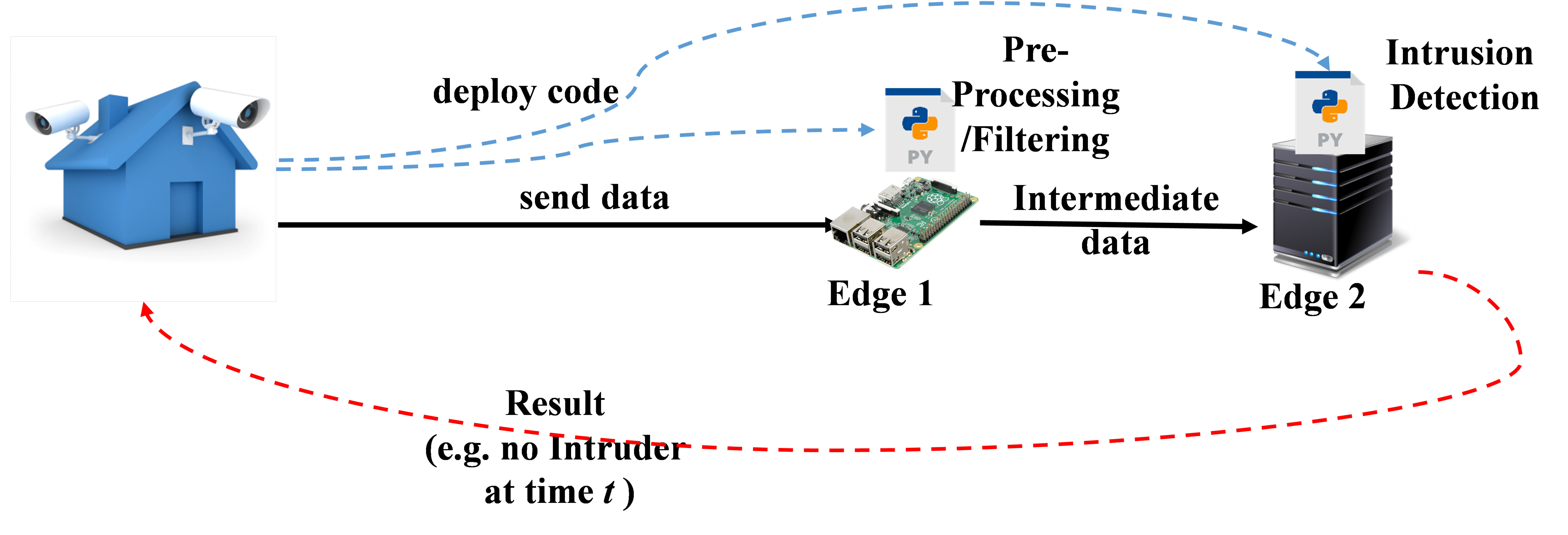}
    \caption{Intrusion detection application using security cameras. Example application that benefits from processing in the TEE.}
    \label{fig:usecase}
\end{figure}

Trusted Execution Environment such as Software Guard Extensions (SGX~\cite{sgx}) is available on Intel processors starting with Skylake. TEE provides a secure area in the processor that protects code and data from all other privileged software on the platform. The code in the TEE is executed safely on secret data that nobody outside the TEE can have access to it including the hardware vendor (e.g., Intel). The privacy and integrity of the code/data inside the TEE is enforced by the hardware. TEEs are sometimes called enclaves, and we use both terms interchangeably in this paper. Intel SGX supports remote attestation of the code and data in the TEE~\cite{attestation}. This enables a remote user to verify the trustworthiness of the hardware and the integrity of the TEE contents (i.e., code and data). The size of the memory reserved to the TEE is limited to 128 MB. However, SGX supports {\it paging} in which the rarely used Enclave Page Cache (EPC) pages are evicted to the unprotected main memory, but they remain encrypted to ensure confidentiality.  With paging, applications in TEE can use more than 128 MB at the expense of encrypting and decrypting the evicted pages which poses an additional efficiency challenge.

\subsection{Example Application and Threat Model}
Figure~\ref{fig:usecase} shows an example application of intrusion detection. In this application, the user installs cameras to monitor the area around his house, and he wants to run computer vision software that analyzes the videos from the installed cameras to detect threats like: (1) the presence of intruders in the backyard, (2) car break-ins, (3) package theft. Many users/application developers do not have the expertise/compute resources to deploy such software services in their houses, so they utilize services provided by public cloud providers (e.g., AWS) or edge providers (e.g., VaporIO). However, there are several key requirements that users need to consider in order to trust such services:
\begin{itemize}
    \item {\bf Code Integrity:} Assuming that the user knows the code/deep learning model he/she wants to run on the data, how to ensure that service provider is running the exact code that the user (app developer) submitted remains a concern. Malicious/Compromised service providers can drop the code and return random results or can inject code that makes copies of private user data.
    \item {\bf Data Privacy:} Even if the code remains unchanged by the cloud/edge service provider, users want to ensure that they get the right predictions without their data being revealed to the service provider. User's data can contain private information about the activities they do around the house,  the cars they own, and the guests coming to visit them. Users want to ensure that the data cannot be seen in an unencrypted form except by the software that analyzes the data.
   
\end{itemize}

{\bf Threat Model:} The security objective of \system is to protect the privacy of user inputs (e.g., home camera feed) that are being used by deep learning inference services hosted by third-party edge/cloud providers. There are four parties in our system: (1) The user that owns the video data (e.g., homeowner), (2) The App developer that developed the Deep Neural Network (DNN) inference service, (3) \system framework that deploys the DNN inference service in the edge devices or cloud platform, and (4) The edge/cloud provider that manages the hardware resources that execute the inference on the user's data. The user trusts the app developer in providing a privacy-preserving DNN inference. The user and the App developer do not trust \system framework to deploy the required services in the Intel SGX devices supported by the cloud provider. However, both the user and the App developer have a method, provided by Intel~\cite{attestation}, to perform remote attestation on all the trusted hardware that they rent to ensure that the code has actually been deployed by Serdab. The edge/cloud provider is the adversary in our threat model, and its goal is to steal user inputs but provide correct output. We assume that the edge/cloud provider has no incentive to produce false outputs, but they have incentive to keep the user's data for use in other activities that are not explicitly approved by the user such as Targeted Advertisements\cite{ads}. This is now common that an attacker wants to use data for their economic benefit so they will not damage the system and quietly leak the data. In our setup, the edge/cloud provider administrators are capable of accessing on-cloud software and hardware resources except the SGX's enclaves. The communication channel from the user's cameras to the enclave and between enclaves is protected by TLS or similar secure protocols. We note that although the data transmission between enclaves has to happen through the untrusted hardware, users can attest that each enclave encrypts its output before being transmitted to the next enclave.
We do not consider the SGX side-channel attacks~\cite{side-attack}, which can be prevented~\cite{side-protect}, as well as Denial of Services (DoS). We assume that the DNN models to be deployed in the cloud are trained in a secured environment, the model parameters are not leaked to adversaries during model training, and the user's camera(s) are not compromised by adversaries.

\section{System Overview}\label{sec:system}

\begin{figure}[t]
      \centering
      \includegraphics[width=0.8\linewidth]{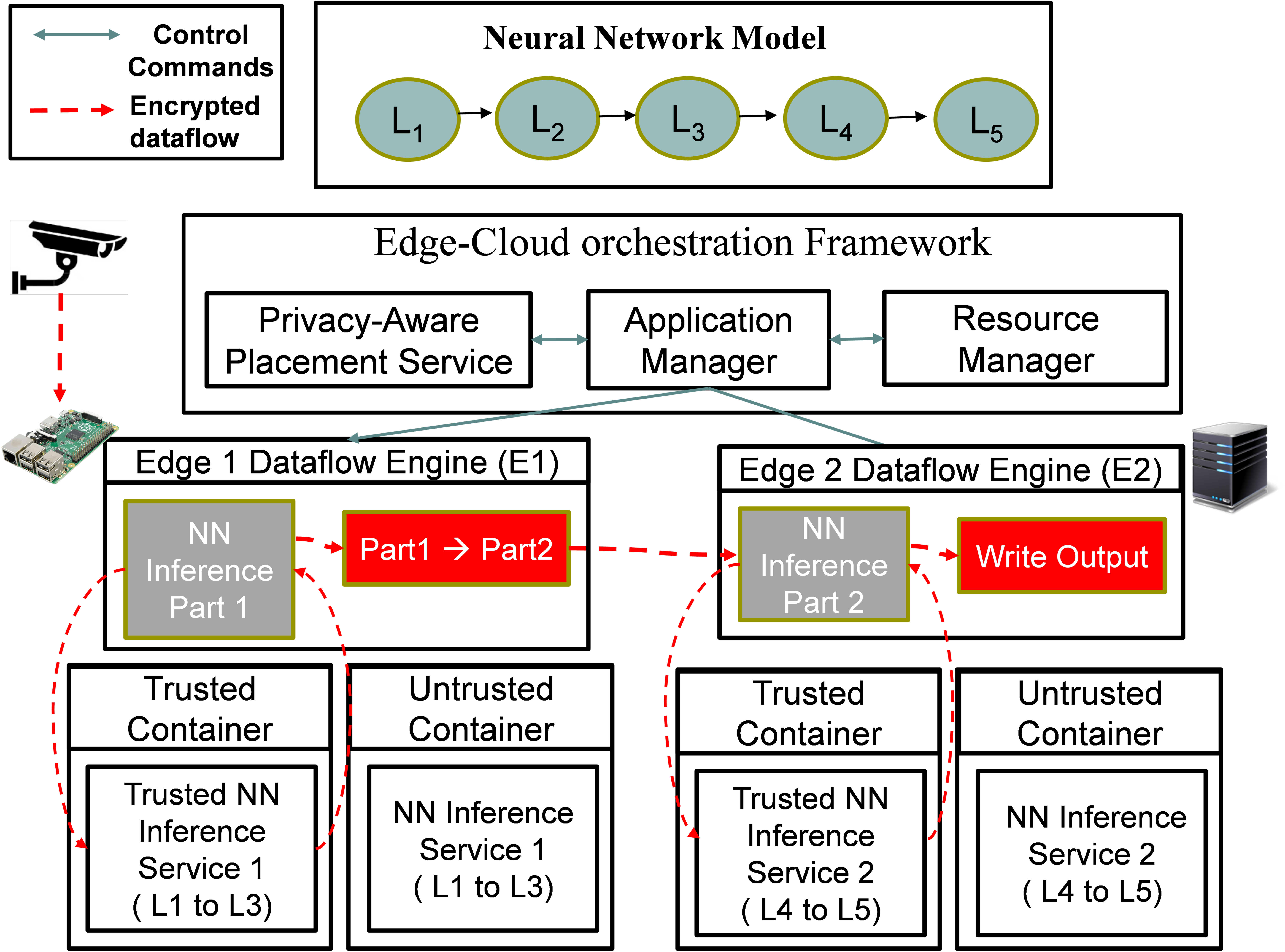}
    \caption{\system System Architecture.}
    \label{fig:system}
\end{figure}

To address the requirements specified in Section~\ref{sec:challenges}, we propose a framework to support distributed analysis of IoT data streams across multiple enclave devices. We focus on visual IoT streams which come from cameras owned by users/organizations (e.g., surveillance cameras, home cameras, wearable cameras, etc). An architectural overview of the proposed framework is presented in Figure~\ref{fig:system}. The top component is the application layer which defines an application in the form of a Directed Acyclic Graph (DAG) of functions/operators. In this paper, we focus on DNN applications in which the application is a NN model, defined as a DAG of NN layers $\{L_x\}_{x=1}^{x=M}$. Each layer $L_x$ represents a compute operation. The layers in a DNN model include: convolutional layers (that combine nearby pixels via convolution operators), pooling layers (that reduce the dimensionality of the subsequent layers), rectified linear unit (ReLU) layers (that perform a non-linear transformation), and fully connected layers (that perform matrix addition, multiplications).  The framework manages multiple devices that could be connected either via local area network or a wide area network. In Figure~\ref{fig:system}, we show two devices (edge 1 and edge 2) at different locations connected via a wide area network and each device has a trusted enclave.

{\bf Edge-Cloud Orchestration:} The second component from the top in Figure~\ref{fig:system} is an edge-cloud orchestration framework which provides an abstraction for deployment and management of NN layers across edge 1 and edge 2 dataflow engines. The orchestration engine is a control component that can be deployed in edge 1 or edge 2 or another device controlled by the user but connected to edge 1 and edge 2 via a local or wide area network. The orchestration engine has a {\it Resource Manager} which carries information about which compute resources are available to execute the NN model (e.g., edge 1 and edge 2). We assume that the edge/cloud provider reports the available resources (devices) correctly. The Resource Manager receives requests to dynamically register new resources and removes old ones. The orchestration framework includes an {\it Application Manager} that sends a request to the local dataflow engine of both edge 1 and edge 2 to deploy the subset of the layers assigned to each device. The {\it Application Manager} also allows encrypted data to flow from one Data-flow Engine to another by deploying a {\it transmission operator} in each Data-flow Engine. The transmission operator pushes the output of the final layer in one Data-flow Engine to the first layer in the next data-flow engine.



{\bf Privacy-aware Placement:} Before the Application manager enacts the deployment, it consults the privacy-aware placement service to find the best placement of layers across devices. The main goal of the privacy-aware placement is to consider the available resources (i.e., edge/cloud devices) to improve the latency of the application without violating the data privacy. 
The privacy-aware placement tries to reduce the amount of layers computed with one TEE through offloading the rest of the layers to another untrusted device or another enclave. However, the privacy-aware placement has to ensure that the output of the layers coming out of one enclave to an untrusted device is {\it sufficiently dissimilar} to the original image. We provide more details about the privacy-aware placement and our definition of image similarity in Section~\ref{sec:approach}.


{\bf Dataflow Engine:} Below the edge-cloud orchestration workflow is the set of dataflow engines. Each edge device has a local stream processing engine that handles the execution of the sub-DAG of layers that are deployed in it.
The dataflow engine plays a management role within one device, particularly it handles the dataflow between the trusted and untrusted hardware in the same device. Each operator in the dataflow engine acts as a client to a service that is implemented in the trusted hardware or a regular CPU. The operator calls the NN Inference service and passes the encrypted data to it. The NN inference will be described in detail in the next section. The calling operator gets back the encrypted result to forward to the next operator. Some operators are transmission operators that do not talk to any service but instead they forward the data over the wide area network to the data-flow engine in another edge device.

{\bf NN Inference Service:} The NN inference service is implemented as a gRPC service enclosed within a Docker container. When the service is initially deployed, \system informs the user to upload the encrypted model parameters directly to the enclave service. The encrypted model parameters will only contain the layers that this enclave is supposed to serve. Once the model parameters are loaded, the gRPC service starts receiving encrypted video frames and executes the inference through Tensorflow Lite (TFlite) library which is a lightweight implementation of TensorFlow~\cite{tensorflow} for resource-constrained devices such as SGX.


\section{Models and Problem Definition}\label{sec:models}
\begin{figure}[t]
      \centering
      \includegraphics[width=0.6\linewidth]{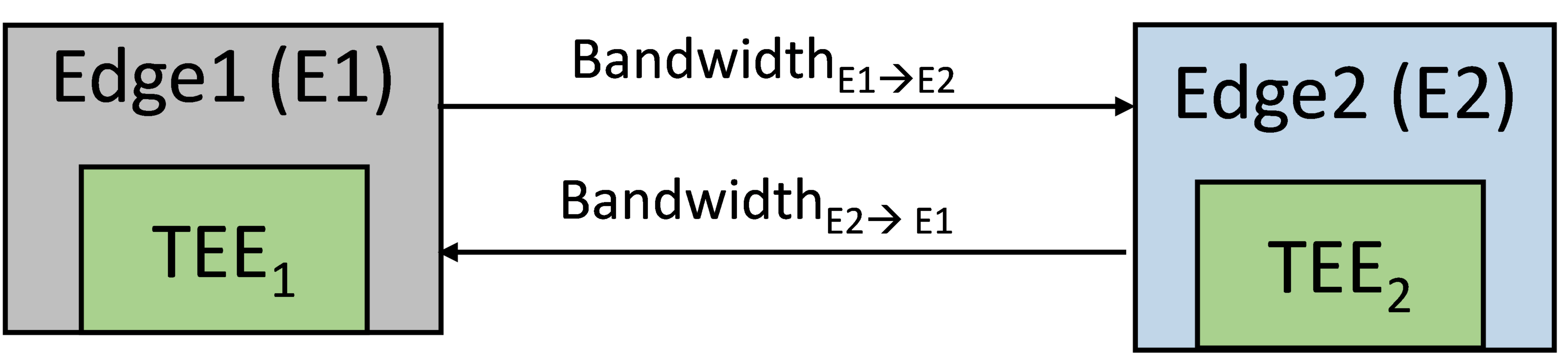}
    \caption{Example Resource graph}
    \label{fig:graphs}
\end{figure}

The performance and memory limitations of TEE (e.g., 128 MB in SGX) make it in-adequate to efficiently support running an entire deep and complex neural network model. Hence, privacy-aware placement aims to explore different ways in which some layers of the neural network are offloaded from a TEE to another device or another TEE with the aim to improve latency while ensuring that the data remains private. In this section, we define the problem of privacy-aware placement. 
We start by describing the NN model, compute resource model, and IoT data model. Then we formally define the problem. In Section~\ref{sec:approach}, we present the intuition behind our approach and the algorithm to solve it. In both sections, we follow the notation in Table~\ref{tab:notation}.

\begin{table}[t]\caption{Table of Notation}
\begin{center}
\begin{tabular}{r c p{5cm} }
$M$ & $=$ & Number of layers in a neural network\\
$NN$ & $=$ & Neural network model with $M$ layers\\
$TEE_l$ & $=$ & Trusted compute Resource\\
$E_l$ & $=$ & Untrusted compute Resource \\
$V_R$ & $=$ & All Compute resources (e.g., $TEE_1$, $E_1$)\\
$V_{R_T}$ & $=$ & Trusted resources (e.g., $TEE_1$, $TEE_2$)\\
$V_{R_{UT}}$ & $=$ & Untrusted resources (e.g., $E_1$, $E_2$\\

$L_x$ & $=$ & Layer $x$ in a neural network\\
$e_{x,E_1}$ & $=$ & Execution time of $L_x$ in $E_1$ \\
$n$ & $=$ & Total number of frames in a chunk \\
$L_x@E_1$ & $=$ & Placement of layer $L_x$ in $E_1$ \\
$L_x \rightarrow L_z@E_1$ & $=$ & Placement of layer $L_x$ to $L_z$ in $E_1$ \\
$p_x$ & $=$ & Resource (device) that $L_x$ is placed on\\
$P_j$ & $=$ & defines a placement path $j$ for each layer in $NN$ to a resource in $V_R$\\

$L_x$ & $=$ & Layer $x$ in a neural network\\
$t_{chunk}\big(n, P_j \big)$ & $=$ & Completion time of a chunk of $n$ frames  given placement $P_j$ \\
$f_y$ & $=$ & Frame $y$ in a chunk of frames \\
$I(L_{x})_y$  & $=$ &   Input to $L_x$ when the input image is $f_y$ \\
$Sim(I(L_1), I(L_x))$  & $=$ &   Similarity bet. input to $L_x$ and input to $L_1$ \\
$\delta$  & $=$ &  Threshold on the similarity between 2 layers \\

$D_{L_x}$ & $=$ & Size (bytes) of output tensor of layer $L_x$ \\
$B_{E_1,E_2}$ & $=$ & Bandwidth (bytes/sec) bet. $E_1$ and $E_2$ \\

$tr{(E_1\xrightarrow{\text{$D_{L_x}$}}E_2)}$ & $=$ & Transmission time (sec) bet. $E_1$ and $E_2$

\end{tabular}
\end{center}
\label{tab:notation}
\vspace{-0.3in}

\end{table}

\noindent {\bf Resource Model:} We model the physical resources as a weighted directed graph $G_R=(V_R,E_R)$, as shown in Figure~\ref{fig:graphs}, where the vertices $V_R = \{E_1,E_2, TEE_1, TEE_2\}$ represent two edge devices $E_1$ and $E_2$. Each device can have a trusted execution environment (TEE). We refer to the TEE inside $E_1$ as $TEE_1$, similarly we refer to the TEE inside $E_2$ as $TEE_2$. We divide the resources (devices) into two distinct sets: trusted resources $V_{R_T}$ and untrusted resources $V_{R_{UT}}$, where $V_{R_{T}} = \{TEE_1,TEE_2\}$, and $V_{R_{UT}} = \{E_1,E_2\}$. The links $E_R=\{(B_{E_1 \rightarrow  E_2},B_{E_2 \rightarrow E_1}) \}$ represent bandwidth availability from $E_1$ to $E_2$ and vice versa.



{\bf IoT Data Model:}
We model the data as an unbounded stream of video frames $f_y$ that can come from one or more camera sources. We aggregate a sequence of video frames into {\it chunks}, where each chunk has duration $T$. The $k^{th}$ chunk $chunk_k$ can be defined as $chunk_k= \langle f_{1}^{k}, f_{2}^{k}, \dots f_{n}^{k}\rangle$, where $n$ is the chunk size. The chunk duration/size is an application-defined parameter to define how often the partitioning algorithm gets invoked to dynamically change the partitioning.



\noindent {\bf Application Model:} As described in Figure~\ref{fig:system}, we model the application as a NN model that has multiple layers $NN=\{L_x\}_{x=1}^{x=M}$. Each layer is a processing element such as convolution or matrix multiplication. A link from layer $L_1$ to layer $L_2$ means that for a given video frame $f_y$, $L_1$ has to be applied before $L_2$ and the output of $L_1$ is the input for $L_2$.

\noindent {\bf NN Layer Profile:}
Each layer in the NN is associated with a {\it profile} which includes:
\begin{enumerate}
\item The cost of executing layer $L_x$ on a video frame $f_y$, when placed on each compute resource. Layer $L_x$ can be placed on $E_1$, $TEE_1$, $E_2$, or $TEE_2$. Their corresponding execution times are $e_{x,E_1}$, $e_{x,TEE_1}$, $e_{x,E_2}$, and $e_{x,E_2}$. The execution includes the time to encrypt the layer's output.

\item The size of output data of layer $L_x$. We denote it as $D_{L_x}$ bytes. The output data of each layer is typically n-dimensional matrix (i.e., tensor) and its size can be derived from the resolution of the input frame.

\item The transmission time of $D_{L_x}$ from $E_1$ to $E_2$, $tr(E_1\xrightarrow{\text{$D_{L_x}$}}E_2) = D_{L_x} / B_{E_1,E_2}  $, where $B_{E_1,E_2}$ is the bandwidth (bytes/sec) from $E_1$ to $E_2$.


\item The similarity between the input to layer $L_x$, denoted as $I(L_x)$, and the original image $f$, which is also the input to the first layer  $I(L_1)$. The similarity is defined by a similarity function $Sim(I(L_1), I(L_x))$ (e.g., Pearson correlation coefficient, mean-squared error). This similarity metric represents how much a layer leaks information about the original image. To compute the similarity, we use a dataset of 1000 diverse images and we get the intermediate output of each layer. For each image $f_y$, we run similarity function between $I(L_1)_y$ and $I(L_{x})_y$, where $I(L_{x})_y$ is the input to $L_x$ when the input image is $f_y$. We compute the overall similarity of the layer as the maximum across all the images:
$Sim(I(L_1), I(L_x)) = max_y(Sim(f_y, {I(L_x)}_y)) $.

\end{enumerate}

{\bf Problem Definition:}
Consider a NN model $NN=\{L_x\}_{x=1}^{x=M}$, which consists of $M$ layers. Let $V_R= V_{R_{T}} \cup V_{R_{UT}}$ be the list of available resources to execute the layers including the list of trusted resources $V_{R_T}$ and untrusted resources $V_{R_{UT}}$. Let $p_{x}$ denote the placement of an arbitrary layer $L_x$ on the compute resources that it will be executed on $p_x = r_k \mid r_{k} \in V_R$. A placement path $P_j=(p_1, p_2,..., p_M)$ defines a placement for each layer in $NN$ to a resource (device) in $V_R$. We note that $P_j$ defines one placement out of combinatorial number of placement of layers in $NN$ on resources $V_R$. For example, a simple placement  $P_1=(TEE_1, TEE_1, ... TEE_1, TEE_1)$ denotes that all the layers are placed on $TEE_1$ while $P_2=(TEE_1, , ..., TEE_2, TEE_2)$ denotes that some layers are placed on $TEE_1$ and some are placed on $TEE_2$. Let $n$ denote the number of video frames (i.e., chunk size) processed by $NN$ and let $t_{chunk}(n, P_j)$ denote the time it takes to perform a complete execution of $NN$ on $n$ video frames given a placement path $P_j$. The goal of the privacy-aware placement is to find the path $P_j^*$ that minimizes the execution time $t_{chunk}(n, P_j)$:

$\quad\quad P_j^* = \underset{P_j} {\mathrm{argmin}} \big(t_{chunk}\big(n,P_j\big)\big)$


while satisfying at least one of the following privacy constraints:


{\bf C1}: Each layer is executed within a trusted device.

$$ \forall p_x \in P_j\ ,\\  p_x \in V_{R_T}$$



{\bf C2:} If a layer is executed in an untrusted device, then the input to that layer has to be sufficiently dissimilar to the input of the first layer of the original image ($I(L_1)$). The similarity is defined by the similarity function and a threshold $\delta$.  

$$ \forall p_x \in P_j\  if\ p_{x} \in V_{R_{UT}},\  then\ Sim [I(L_x),I(L_1)] < \delta $$



\section{Privacy-Aware Placement}\label{sec:approach}

\begin{figure}
    \begin{minipage}{.15\textwidth}
        \includegraphics[width=\linewidth]{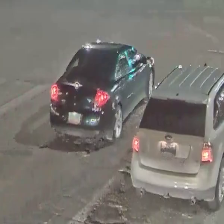}
    \end{minipage}
    \begin{minipage}{0.15\textwidth}
         \includegraphics[width=\linewidth]{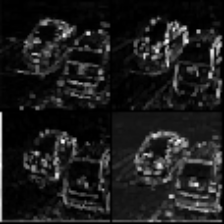}
    \end{minipage}
    \begin{minipage}{0.15\textwidth}
         \includegraphics[width=\linewidth]{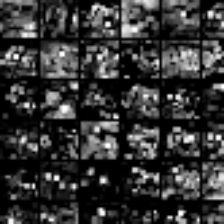}
    \end{minipage}
\caption{Output of intermediate layers of GoogLenet: left (original image), middle (layer 1: maxpool), right (layer 4: maxpool) }
\label{fig:layers}
\end{figure}

\begin{figure}[t]
      \centering
      \includegraphics[width=0.8\linewidth]{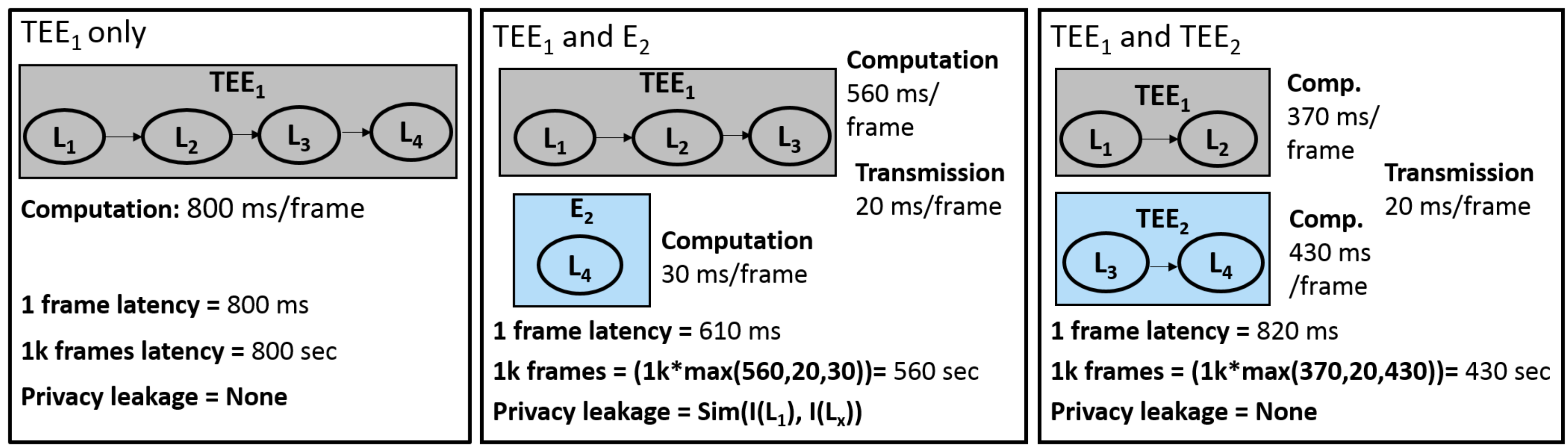}
    \caption{Comparison between the execution time of a NN in three different cases: all layers in $TEE_1$ (left), layers partitioned across $TEE_1$ and $E_2$ (middle), layers partitioned across $TEE_1$ and $TEE_2$ (right).}
    \label{fig:example}
\end{figure}

Our privacy-aware placement technique explores alternative placement that can potentially improve the overall latency of the NN application without violating the privacy. Our privacy-aware placement method relies on two {\bf key insights}:
\begin{itemize}
\item {\bf dissimilar intermediate outputs:} Several studies have been conducted to understand the internal mechanisms of DNNs~\cite{vis}. Such studies have shown that for an image classification DNN, the first few layers (shallow layers) represent low-level image processing operations such as edges, corners, and contours of the original inputs.
On the other hand, later layers (i.e., deep layers) represent more abstract and class-specific information related to the final outputs. Figure~\ref{fig:layers} shows an illustrative example for the output of layer 1 and layer 4 of Google's InceptionNet (GoogLenet). As shown in Figure~\ref{fig:layers}, layer 1 does primitive image processing operations (e.g., edge detection) so the content of the image is still visually identifiable by humans acting as adversaries. However, the output of layer 4 is significantly less similar to the original image. This insight can be used in the context of TEE to execute the first 4 layers inside the enclave until the image becomes dissimilar to the original image. The rest of the layers (after layer 4) can then be executed in a regular processor to leverage hardware accelerators such as GPUs. However, the question becomes what is the minimum number of layers to be executed within the resource-constrained TEE that is enough to conceal the identity of the data. To answer this question, we measure the correlation between the original input image and each layer's output.
After experimenting with several correlation metrics (e.g., mean-squared-error (MSE), Pearson correlation coefficient, and structural image similarity (SSIM)) and conducting a user study, we realize that the most crucial metric that affects the correlation between the input image and the intermediate layer is the \emph{resolution} (i.e., number of pixels) of intermediate layer's output. We notice that the output of each layer has a grid of images (Figure~\ref{fig:layers}), where each image in the grid is either smaller than or equal to the images in the input grid. This happens because convolution and pooling operations reduce the number of pixels in each image in the grid. We observe that when the output of a layer has images
with resolution less than or equal a certain threshold (e.g., 20x20 pixels) then such output has undergone enough transformations in which it cannot be visually identified no matter how much you can resize it. Hence, we conclude that the output of layer $L_x$ is considered private if the resolution of its output is below certain threshold $\delta$. We note that our privacy-placement method is not restricted to using the resolution as a metric and more complex similarity metrics can be utilized.

\item {\bf pipeline parallelism:} A key idea in our privacy-aware placement approach is to allow processing a stream of video frames in a pipeline fashion. Pipelining allows utilizing both $TEE_1$ and $TEE_2$. For example, while the $TEE_1$ is processing the first part of the neural network for the second video frame, $TEE_2$ will be processing the second part of the neural network for the previous frame. Figure~\ref{fig:example} shows the execution and transmission time for executing a neural network in three different cases: (1) all layers of the NN are deployed on the $TEE_1$ (left column), (2) the NN is partitioned across $TEE_1$ and $E_2$ (middle column), and (3) the NN is partitioned across $TEE_1$ and $TEE_2$ (right column). Figure~\ref{fig:example} shows that the best completion time for one frame is the second case when NN is partitioned across $TEE_2$ and $E_2$ (610 ms). However, when we have a stream of 1000 frames, the best completion time is the third case (i.e., Multiple TEEs). The reason is \emph{pipelined execution} which means that when both $TEE_1$ and $TEE_2$ are concurrently processing different frames, the completion time becomes bounded by the completion time of the slowest device. In the case of one enclave and one regular CPU (middle column), the enclave has to process more than half of the layers to reach a point where the resolution is less than threshold $\delta$, which makes the bulk of the workload skewed towards the slower device ($TEE_1$). However, in case of multiple TEEs, the layers can be more evenly divided across the $TEEs$ which results in better chunk completion time than the other two cases. Using multiple enclaves has an additional benefit that it has no privacy leakage because the intermediate data can only be decrypted in $TEE_2$ unlike the middle case in which the intermediate data can be processed in untrusted processor $E_2$.
\end{itemize}

\begin{figure}[t]
      \centering
      \includegraphics[width=\linewidth]{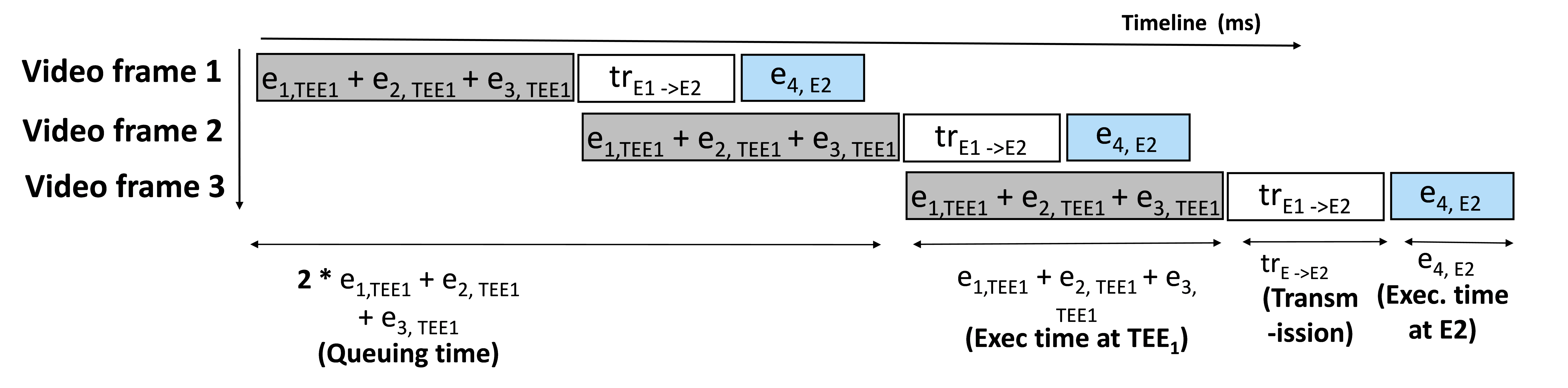}
    \caption{Pipelined execution of 3 video frames ($n=3$) for the partitioned NN in the middle column of Figure~\ref{fig:example}.}
    \label{fig:pipeline}
\end{figure}

{\bf Cost Calculation for different placements:} To better illustrate how the completion time of a partitioned NN is calculated, we show an example of the pipelined execution in Figure~\ref{fig:pipeline}. Figure~\ref{fig:pipeline} depicts the execution for a {\it chunk} of three video frames.
From Figure~\ref{fig:pipeline}, we notice that at the same time that video frame 2 is being processed at $TEE_2$, video frame 1 is being transmitted over the network and processed at $E_2$. Hence, the completion time of executing a chunk of $n$ frames on a neural network that follows placement $P_j= (TEE_1, TEE_1, TEE_1, E_2)$ is calculated as:

\begin{equation}\label{eq:pipelining-example1}
\begin{split}
t_{chunk}\big(n=3, P_j) = 3*(e_{1, TEE_1} + e_{2, TEE_1}  \\ + e_{3, TEE_1})  + tr(E_1\xrightarrow{\text{$D_{L_1}$}}E_2)  + e_{4, E_2}
\end{split}
\end{equation}

We note that for a large chunk of frames $n$, the chunk completion time for the example in Figure~\ref{fig:pipeline} becomes bounded by the first term only, which is the queuing time in the slowest device $TEE_1$.

\begin{equation}\label{eq:pipelining-example2}
\begin{split}
t_{chunk}\big(n, P_j) = n*(e_{1, TEE_1} +   e_{2, TEE_1} \\ + e_{3, TEE_1})  + tr(E_1\xrightarrow{\text{$D_{L_1}$}}E_2)  + e_{4, E_2} \\  \simeq n*(e_{1, TEE_1} + e_{2, TEE_1} + e_{3, TEE_1})
\end{split}
\end{equation}
\vspace{-0.1in}




\begin{figure}[t]
      \centering
      \includegraphics[width=0.6\linewidth]{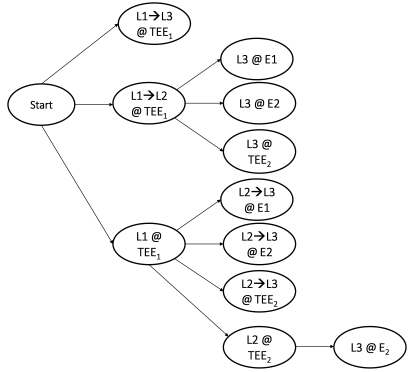}
    \caption{Placement Tree that shows possible placements for the example in Figure~\ref{fig:pipeline}. Each path is a solution to the privacy-aware placement problem}
    \label{fig:placement}
\end{figure}

{\bf Algorithm Steps:} The following steps are performed offline once to find the best initial placement, then, the system keeps monitoring the online profiling information for the execution time of each NN layer and issues a re-partitioning when the profiling information deviates from the predicted execution times. 
\begin{enumerate}
   
    \item {\bf Step 1 - Construct placement tree to explore possible partitioning points}: Figure~\ref{fig:placement} shows an example placement tree for a NN with 3 layers. Each node in the tree shows the placement of one or several layers, where $L_1 @ TEE_1$ denotes that $L_1$ is placed on $TEE_1$ while $L_1 \rightarrow L_3 @ TEE_1$ denotes that $L_1$, $L_2$, $L_3$ are placed on $TEE_1$. Each path $P_j$ in the placement tree is a possible solution for the privacy-aware placement problem. The first level starts with possible placements in $TEE_1$ because the processing has to start in a trusted resource. The second level shows possible placements of the second part of the NN on either $E_1$, $E_2$, or $TEE_2$ and the third level of the tree shows the possibility of offloading layers from $TEE_2$ to $E_2$. Another level can be added to the tree if the application requires that the final layers get executed where the processing started in $TEE_1$. This is beneficial if the application requires the object labels to be generated in a trusted resource.
    
    \item {\bf Step 2 - Evaluate the execution time for each path in the placement tree}: For each path $P_j$ in the placement tree, we compute two values: (1) the completion time $t_{chunk}(n,P_j)$ according to Equation~\ref{eq:pipelining-example2}, (2) the maximum similarity  computed as:
        $Sim_{P_j} = max(Sim(I(L_1), I(L_x)), \forall p_x \in P_j\ and\ p_x \in V_{R_{UT}}$
        
    This value defines the maximum privacy leakage across the placement path.
 
    \item {\bf Step 3 - Choose the optimal placement path}: From step 3, we get two sets $S_{completion}$ and $S_{Sim}$. The first set contains completion times and second set contains the similarity (i.e., privacy) value for each path $P_j$ out of the $N$ possible paths in Figure~\ref{fig:placement}:
    
        $$S_{completion} = \{t_{chunk}\big(n,P_j\big) \}_{j=0}^{j=N}$$
        $$S_{Sim} = \{Sim_{P_j}\}_{j=0}^{j=N}$$

    We choose the optimal placement $P_j$ that yields the minimum completion time while the similarity is less than threshold $\delta$:
    
    $$ j^*= \underset{j} {\mathrm{argmin}}\big(S_{completion} \big)  \quad such\ that\ Sim_{P_j} < \delta$$

    
\end{enumerate}

{\bf Algorithm analysis:} The algorithm complexity is bounded by the number of paths $N$ in the placement tree. As shown in Figure~\ref{fig:placement}, the tree has 3 levels. To calculate the complexity, we compute the maximum degree at each level of the tree. We denote the maximum degree at level 1 as $deg_1$. The maximum degrees at levels 2 and 3 are denoted as $deg_2$ and $deg_3$, respectively. The upper bound on the total number of paths $N=O(deg_1 * deg_2 * deg_3)$. Level 1 has $M$ nodes because there are $M$ possible ways to partition a NN of $M$ layers across $TEE_1$ and another device, therefore $deg_1=M$. Each node in level 2 denotes where the second part of the NN is executed. The second part can be executed in $E_1$, $E_2$, or distributed among $TEE_2$ and $E_2$. Since there are $M-1$ ways to distribute $M-1$ layers across $TEE_2$ and $E_2$, then $deg_2 = 2 + (M-1) = M + 1$. Level 3 will only have one possibility which is deploying the rest of the layers on $E_2$ so $deg_3=1$. Based on the previous analysis $N=O(M * (M+1) * 1 )= O(M^2)$. We note that this analysis is based on the assumption that we have 2 TEEs, in the general case with $R$ TEEs, $N=O(M^R)$, where $R$ is expected to be a small constant significantly lower than the number of layers $R<<M$.   


\section{Evaluation}\label{sec:eval}

{\bf System setup:} The computing infrastructure consists of two desktops. Both desktops have Intel Core i7-9700k (4.6 GHz) CPU and 32 GB of memory. Both desktops are equipped with Intel SGX trusted enclave technology. Each server has an Nvidia RTX 2080 GPU. We control the bandwidth between the two machines to be 30 Mbps which simulates an average wide area network connection. We use Google Asylo~\cite{asylo} to deploy gRPC services in the enclave and the NN inference is done through TFtrusted~\cite{tftrusted-github} which provides integration between Asylo, gRPC and TensorflowLite. Each of the devices has a local dataflow engine, Apache NiFi, that handles data transfer between operators within the device and we use Echo~\cite{echo} orchestration framework to handle the communication between the two Nifi instances.

{\bf Models:}
We evaluate our system with well-known Convolutional neural network (CNN) models: (1) GoogLeNet, (2) Alexnet, (3) Resnet, (4) Mobilenet, (5) Squeezenet. We download the models from the Tensorflow Model Library~\cite{tfmodels}, pre-trained with the Imagenet (ILSVRC 2012) dataset.

{\bf Datasets:} We experiment with three video surveillance datasets~\cite{noscope-github}. The datasets vary in the object types (car, person, boat), and locations (indoor, outdoor, different cities). We experiment with surveillance datasets because they include sensitive content such as faces and car license plates that require processing in a privacy-preserving fashion. Our dataset consists of 1 hour from each surveillance video and we sample one frame per second so we get a total of 10800 frames. The resolution of each frame is restricted to 224x224 pixels because it is the input resolution required by all the models.


 

\subsection{Latency and Resolution of Intermediate Layers Output }

\begin{figure}[t]
      \centering
      \includegraphics[width=0.83\linewidth]{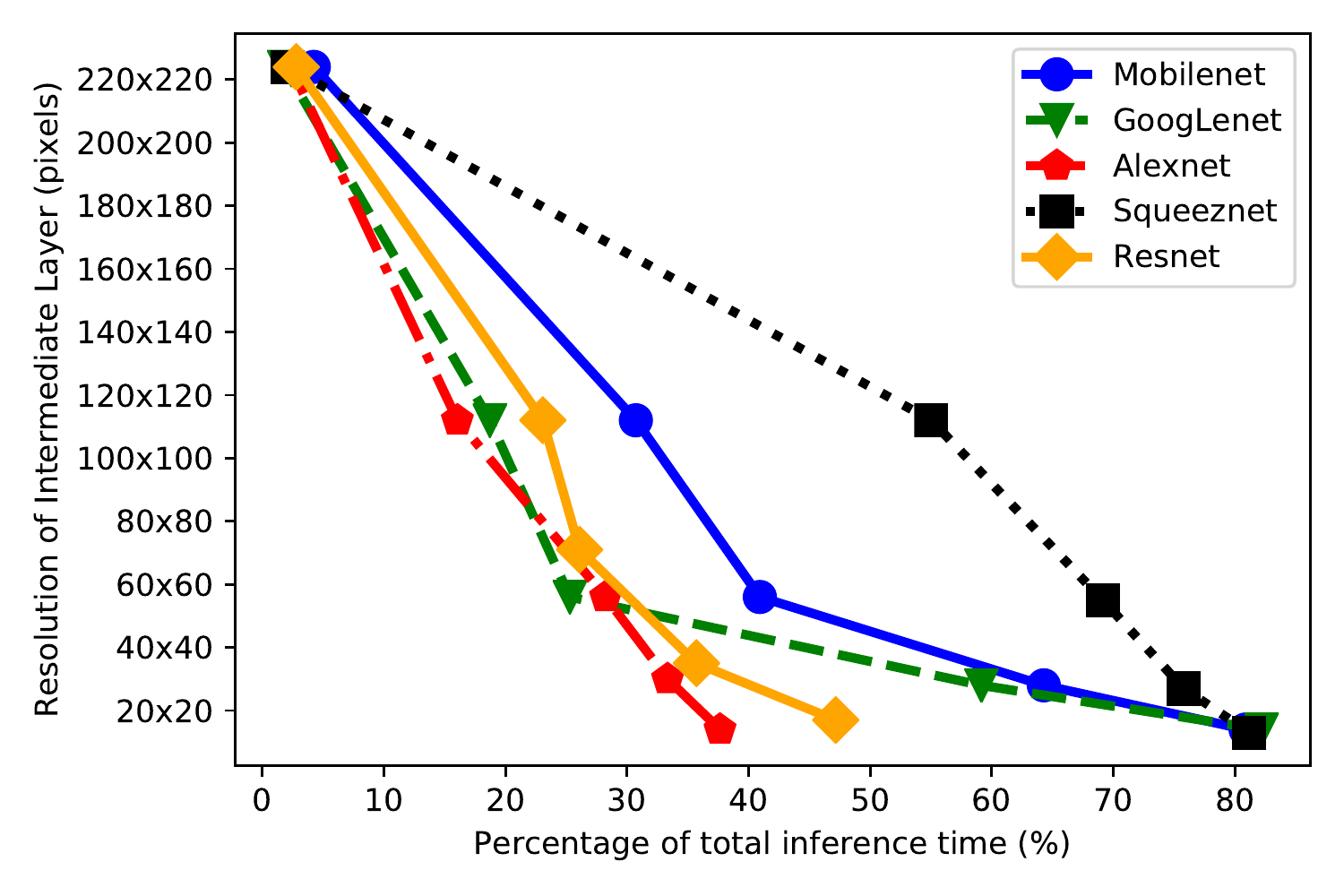}
    \caption{Relationship between the percentage of time spent in executing inference and the resolution of the intermediate output.}
    \label{fig:corr-latency}
\end{figure}

 In this section, we conduct an experiment to show the relationship between the  latency of computing the output of intermediate NN layers and the similarity between the intermediate output compared to the original image. For each of the five NN models, we pass an individual video frame to the NN and we compute the latency of computing each of the intermediate layer outputs. To visualize the intermediate layer, we convert the tensor to a grid of small images (Figure~\ref{fig:layers}) using TensorFlow CNN visualization tool. To assess the similarity between intermediate layers and the original image, we get the resolution of a single image in the grid. 
 The resolution of a single image gives an estimate of the privacy leakage from this intermediate layer. In Figure~\ref{fig:corr-latency}, we plot the relationship between the percentage of time spent in computing the intermediate output vs. the resolution of this intermediate output. Figure~\ref{fig:corr-latency} shows that the deeper the layer is, the more time one spends to compute its output and the less its resolution (i.e., correlation with the original input) will be.
 However, an interesting insight in Figure~\ref{fig:corr-latency} is that different models tend to have different trends. For example, for models like GoogLeNet, Squeezenet, one needs to spend 80\% of the entire inference time to reach an intermediate output with resolution of 20x20 pixels or less. However, Alexnet and Resnet reach such resolution in less than 50\% of the inference time. From this experiment, we conclude that each model can be partitioned differently based on how fast the resolution drops. We will discuss different partitioning strategies in Section~\ref{sec:nn-results}.

\subsection{User Study to find Resolution Threshold}

\begin{figure}[t]
      \centering
      \includegraphics[width=0.9\linewidth]{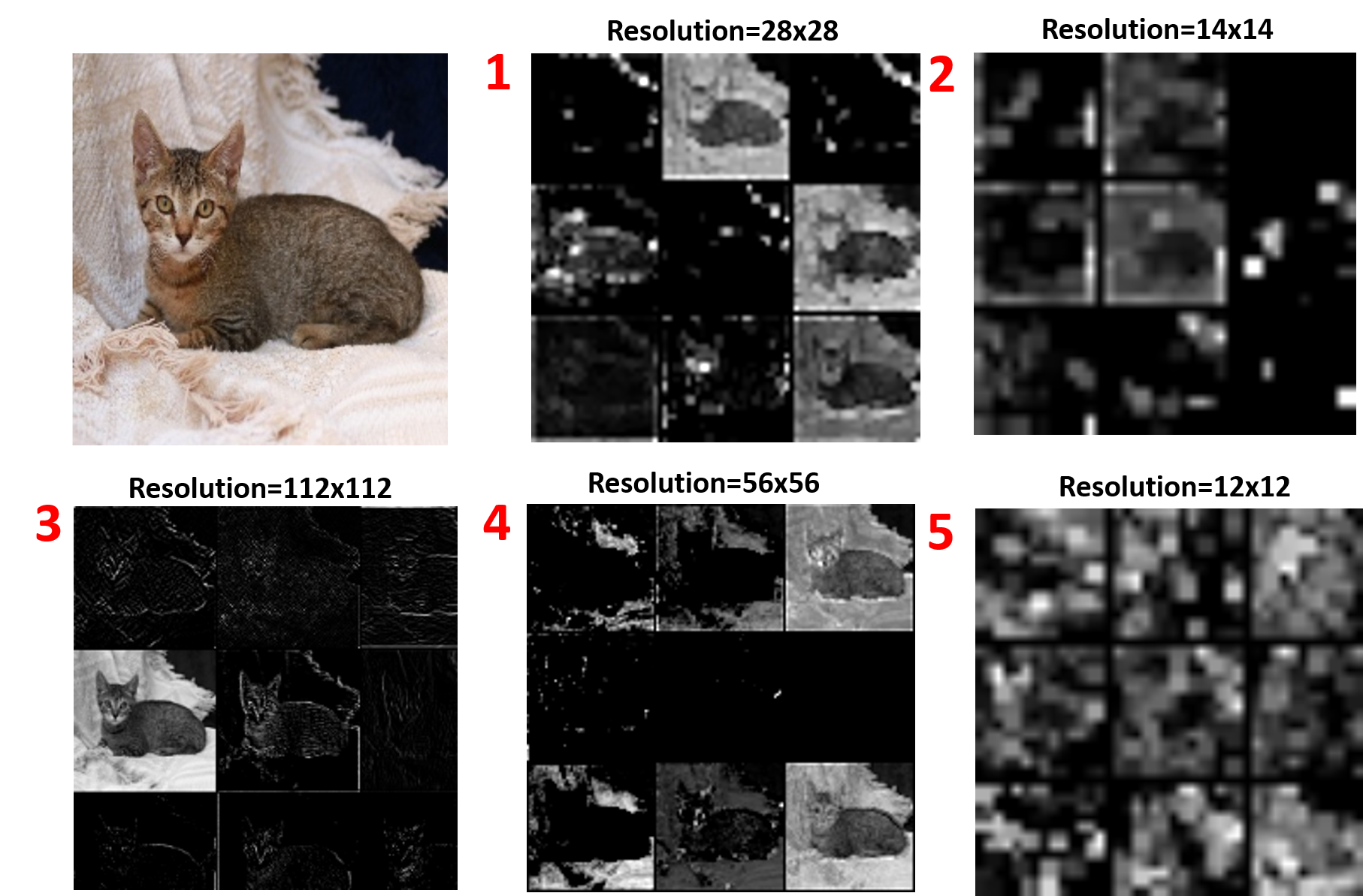}
    \caption{
Example question in the second part of the user survey. Users are asked to rank the images according to the similarity to the original image (Resolution values are not shown to users).}
    \label{fig:ques}
\end{figure}

In this section, we conduct user study to show the relationship between the resolution of the intermediate layer's output and the ability of human subjects to recognize the objects in the image. The study is divided into two parts. In the first part, we display the intermediate output of several layers from each model and we ask each user to identify the object in this intermediate output. In the second part of the survey, we show the users an original image and 5 sample images, where each image is an output of a random layer of the same NN model. The users are asked to rank these images based on how representative they are compared to the original image. Ten subjects took part in this study. They had normal/corrected-to-normal vision.

\begin{figure}
 \centering
    \begin{minipage}{0.24\textwidth}
     \includegraphics[width=\linewidth]{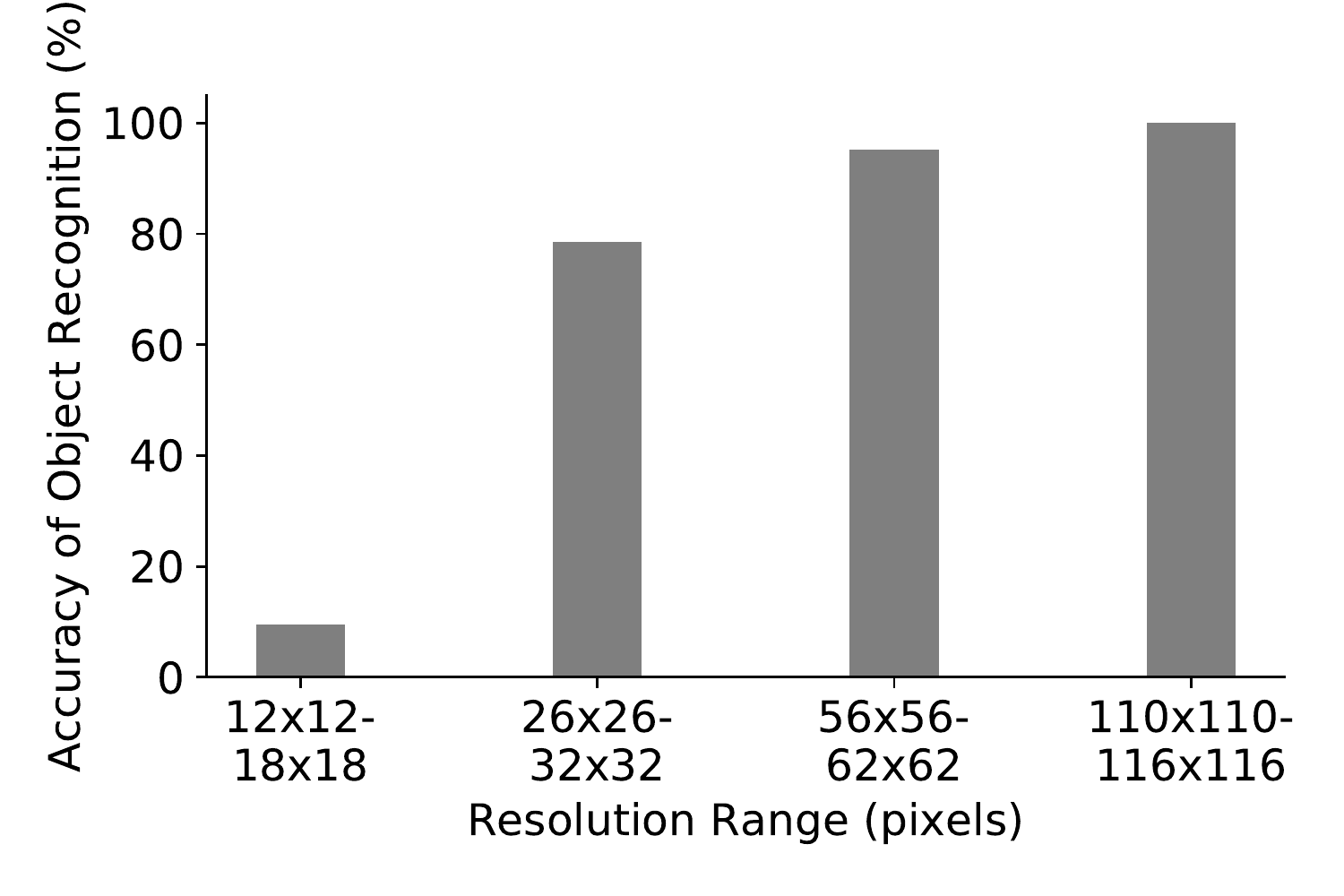}
    \caption{Accuracy of object recognition by human subjects at different resolution ranges.}
    \label{fig:user-study-1}
    \end{minipage}
     \begin{minipage}{0.24\textwidth}
     \includegraphics[width=\linewidth]{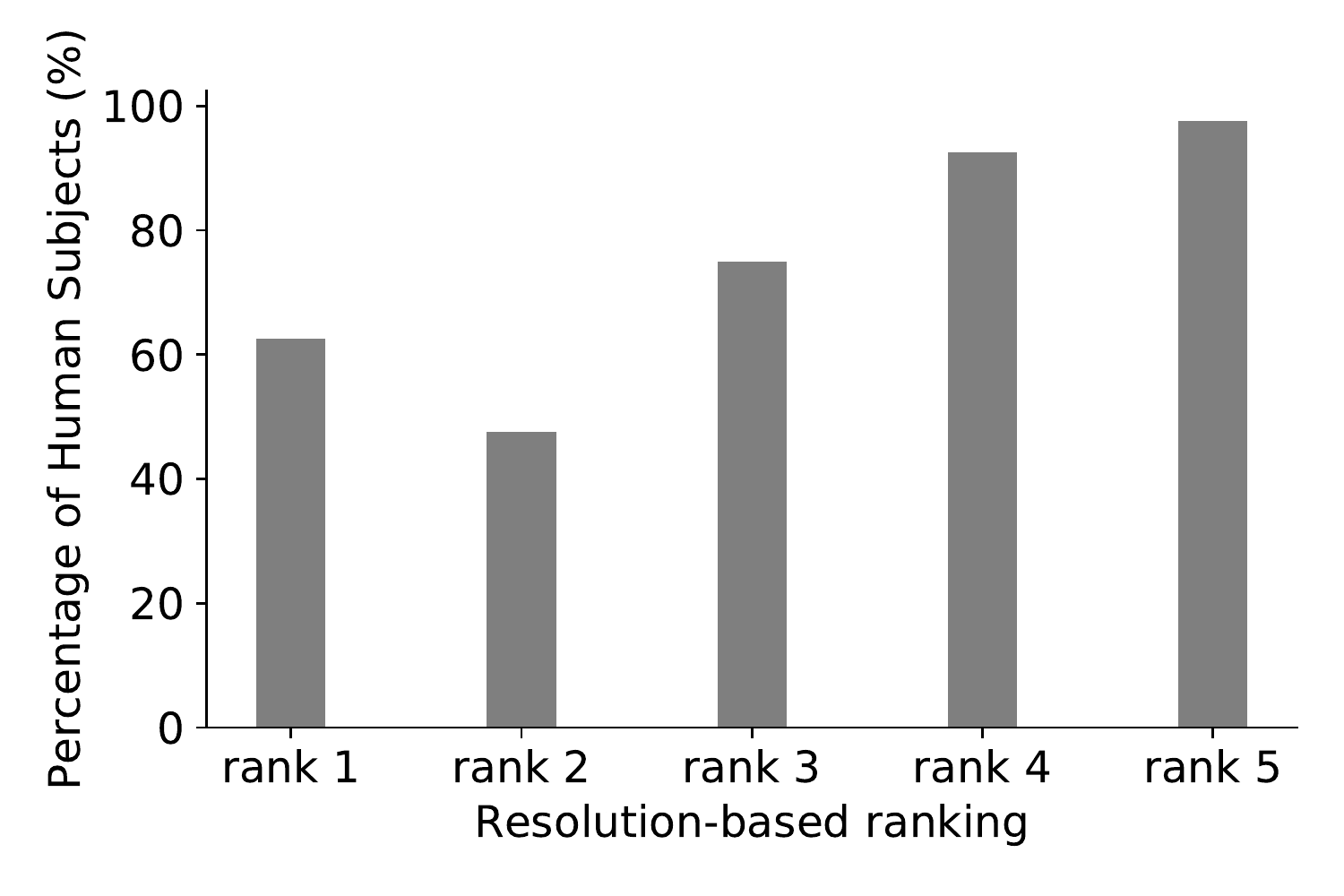}
    \caption{Percentage of human subjects who rank an image in the same position as resolution-based ranking.}
    \label{fig:user-study-2}
    \end{minipage}

\end{figure}

{\bf Data Generation Process:}
For the first part of the study, we collect 100 images from Imagenet's image classification dataset. Each image has a single object and the objects belong to 10 classes: Cat, Dog, Car, Truck, Bus, Aeroplane, Boat, Horse, Elephant, and Person.  We use the same 5 NN models, described in the previous section, to generate the intermediate output. For each model, we pick 5 layers that have distinct resolution values. We choose a random image from the dataset to generate the output of each layer. We end up with 25 outputs for this part of the survey and we ask the users to identify the object in each of the 25 images.
For the second part of the survey, we choose one image at random for each model, then we pass the image by different layers of the model, and we generate the output of 5 layers that have distinct resolution values. We shuffle the order of the layers' outputs and we ask the users to rank them based on the similarity to the original image. Figure~\ref{fig:ques} shows an example question shown to the users.

{\bf Protocol:} Each subject is asked to fill the survey individually. We present each subject with a link to a web form with 30 questions: in 25 questions he/she is asked to identify the object in the image and in 5 of them he/she was asked to sort images based on their similarity to the original image. We ask the subjects to resize the images as much as they can to try to identify the objects.

{\bf Discussion and Results:} In the first part of the survey, we calculate the accuracy at which the users were able to identify the objects. In Figure~\ref{fig:user-study-1}, we plot the accuracy with respect to the resolution values. The figure shows that human subjects were able to identify the object with 100\% accuracy when the resolution is above 110x110. The accuracy degrades slightly when the resolution is in the range 26x26 - 32x32 pixels, and it drops drastically when the resolution is in the range 12x12 - 18x18 pixels. 
Hence, we conclude that a resolution below 20x20 pixels is the sweet spot at which the objects in an image are hardly identifiable. For the second part of the survey, we evaluate how often the ranking of humans matches the ranking based on the highest to lowest resolution. For each question (e.g., Figure~\ref{fig:ques}), we rank the images from 1 to 5 based on the resolution where rank 1 represents the highest resolution and rank 5 represents the lowest resolution. For each ranked image, we calculate the percentage of human subjects who ranked it the same way as the resolution. We plot the results in Figure~\ref{fig:user-study-2}.
The results show an interesting insight that human subjects have different opinions about which images rank as the most similar to the original image (rank 1). On the other hand, there was a general consensus among human subjects and resolution metric about which images should rank last (ranks 4 and 5). We conclude that when the resolution is high, the objects are well identifiable and human subjects rank images differently, but everybody agrees on the ranking when the image resolution is below 20x20 pixels.

\subsection{Neural Network Partitioning}\label{sec:nn-results}
In this section, we show the impact of our neural network partitioning. We compare between the following strategies:
\begin{enumerate}
    \item {\bf 1 TEE:} the entire NN is deployed in one TEE,

    \item  {\bf No pipelining:} this approach is adopted by related work in mobile computing literature~\cite{mcpp} and a recent NN inference system (Neurosurgeon~\cite{neurosurgeon}). In those approaches, NN partitioning is designed to minimize the latency of one frame ($n=1$) and they do not consider the cases when a stream of frames arrives in a short period of time.

    \item  {\bf 1 TEE \& 1 GPU:} We use the proposed method in Section~\ref{sec:approach} to find the layer to offload to the GPU such that the layer we partition at has an output with resolution less than 20x20 pixels (In this approach, we do not consider having the second TEE in the available resources).

    \item  {\bf 2 TEEs:} The neural network is partitioned across two TEEs according to the algorithm in Section~\ref{sec:approach} (In this approach we do not consider offloading to a CPU/GPU).

    \item {\bf Proposed:} We partition the neural network considering all available resources (2 TEEs and one GPU).
\end{enumerate}







\begin{figure}[t]
      \centering
      \includegraphics[width=0.8\linewidth]{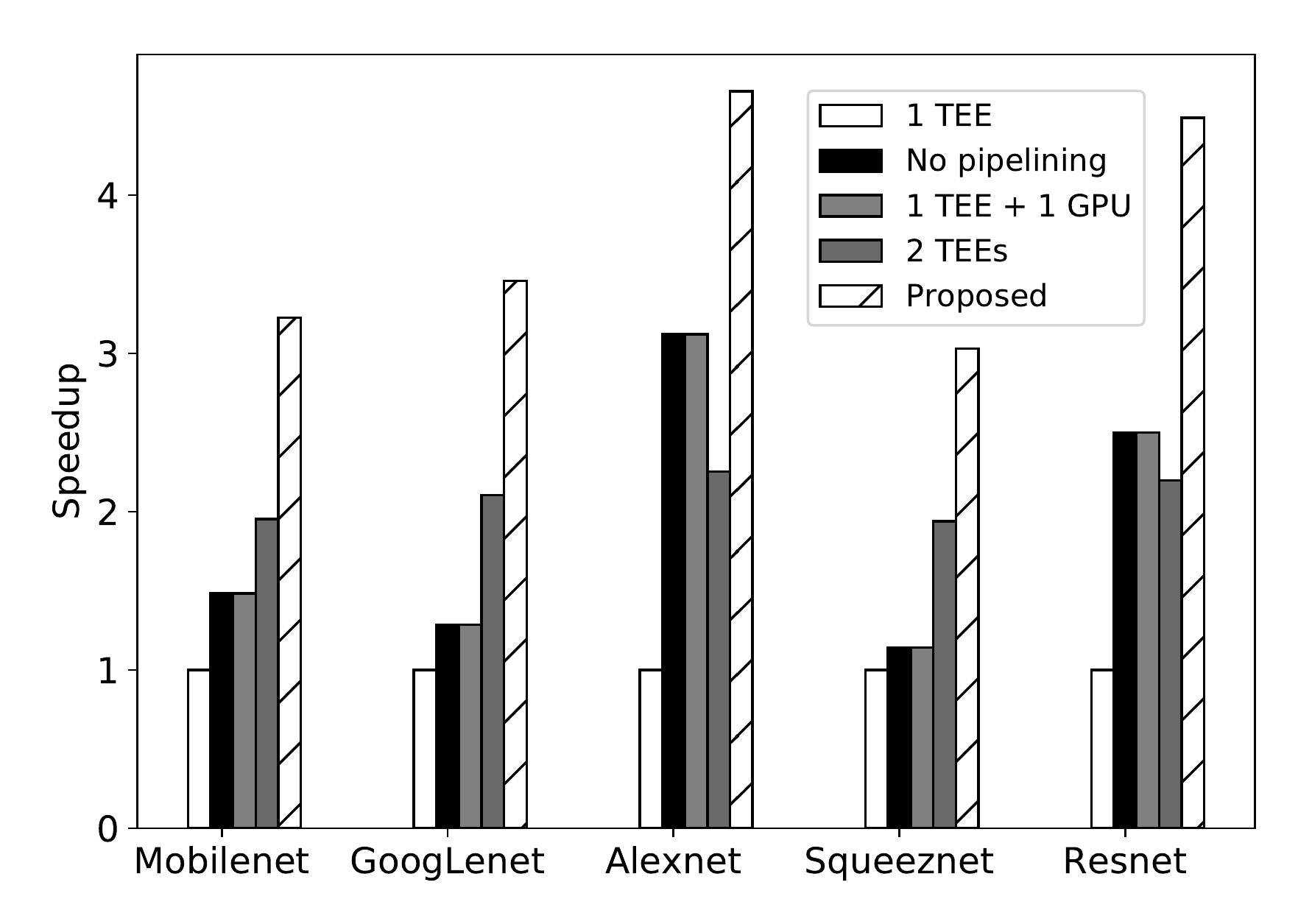}
    \caption{
    Speedup comparison between various NN partitioning approaches.}
    \label{fig:speedup}
\end{figure}

Figure~\ref{fig:speedup} shows the speedup in the end-to-end execution of the entire dataset of 10800 frames for the different approaches compared to the baseline that uses 1 TEE for the entire processing. For the approaches that employ partitioned NN, the speed up is measured on the end-to-end execution time which includes: the time to process both parts of the NN, and the time to encrypt and transmit intermediate outputs. From the results, we observe that for three of the above models (GoogleNet, Mobilenet and Squeezenet) using 2 TEEs outperforms using 1 TEE and 1 GPU because in the latter case most of the processing happen in the slow TEE which results in 1.15-1.5x speedup over using 1 TEE for the entire execution. On the other hand, 2 TEEs can achieve 1.8 to 1.95x speedup for the same three models because the model processing is almost equally distributed across the enclaves. However, the results are different in the other 2 models (Alexnet, Resnet) because the majority of the workload happens in the fast GPU resulting in 2.5-3.1x for 1 TEE \& 1 GPU compared to a speedup of 2.2-2.3x for 2 TEEs.


We note that the proposed approach can go beyond using 2 TEEs by using 2 TEEs and 1 GPU resulting in the best speed up 3.2-4.7x. The best speedup happens in Alexnet because each TEE can do only 19\% of the processing leaving the remaining 62\% to the fast GPU.
Our final observation on Figure~\ref{fig:speedup} is that the {\it No pipelining} baseline ends up choosing the same decision as 1 TEE \& 1 GPU because its partitioning decision is based on one frame only and it fails to consider that the second TEE used to process the next frame while the first TEE is processing the current frame.



\subsection{Inference latency and overhead on TEE}

\begin{figure}[t]
      \centering
      \includegraphics[width=0.85\linewidth]{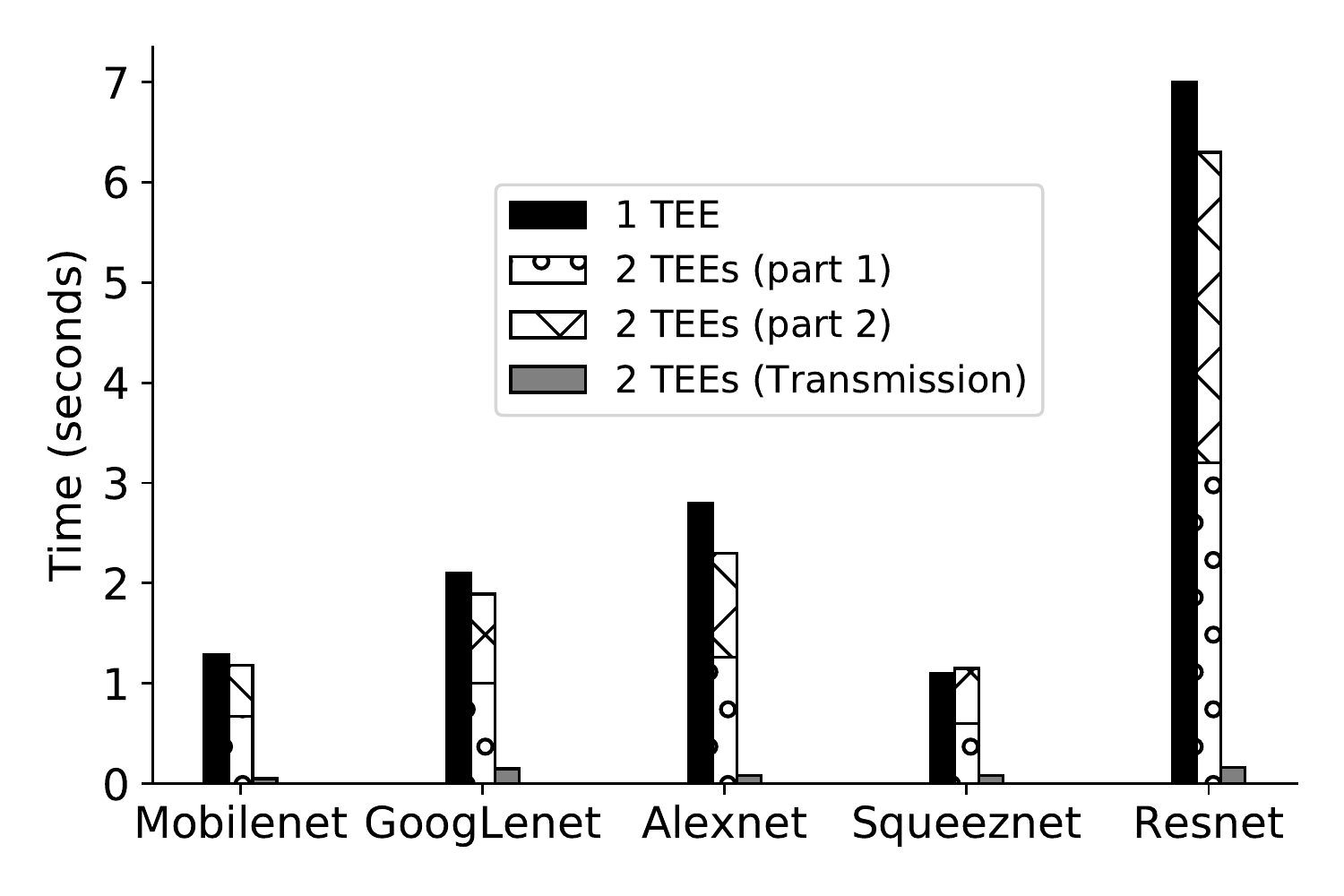}
    \caption{
Execution time of NN inference per frame when deployed in 1 TEE and 2 TEEs.}
    \label{fig:times}
\end{figure}

In this section, we analyze the time it takes to perform an inference on a single frame using one vs. two TEEs. We note that in the case of two enclaves we analyze the time to process the first part of the NN in $TEE_1$, the time to encrypt the intermediate output in $TEE_1$, the time to transmit the intermediate output to $TEE_2$, the time to decrypt the intermediate output, and, the time to process the second part of the NN in $TEE_2$. We note that not all these times are added together in the presence of a stream of frames because of pipeline parallelism but in this section we show it for a single frame. We show a breakdown of the different execution times in Figure~\ref{fig:times}. The figure shows that distributing the layers across the enclaves have an additional benefit that it reduces the memory requirements per enclave resulting in a faster overall execution. As shown in the figure the sum of execution times in both enclaves is less than the execution time of the entire network in one enclave in 4 of the 5 models. The result is more pronounced in Alexnet because it is the largest model (243 MB). Conversely, Squeezenet does not show a similar trend because it is the smallest model (5 MB) and partitioning its layers does not have a significant impact on the memory. We notice that the encryption and decryption times using Advanced Encryption Standard (AES) with 128-bit key is negligible (i.e., less than 2.5 ms/frame) compared to the other times so we omit it from the figure. The transmission time ranges from 0.01 to 0.12 seconds based on the size of the intermediate layer and it is less significant than computation time. The computation time represents the largest portion of the processing due to the limited hardware resources in the TEE, the computation time ranges from 1.1 seconds for Squeezenet to 7.2 seconds for Resnet.

\section{Discussion}\label{sec:discussion}
{\bf Other deep learning models:} The deep learning models we experimented with in this paper belong to the object detection/classification category which represents an important and widely deployed subset of all video processing. However, many of these models produce lower resolution images deeper down the network which is not the case for other deep learning architectures such as UNet architecture which is the state of the art for image segmentation tasks.  For UNet-like architectures, our approach should handle it using the multiple enclaves setup in which each subset of the model is executed inside a trusted enclave device.



{\bf Security Attacks and Guarantees:} Previous research~\cite{reconstruct} has shown some attacks that intend to reconstruct the original input from the intermediate layers output of a NN. Such attacks are also known as {\it Input Reconstruction Attacks}. To perform such attacks, the adversary has to either have (1) the model parameters that lead the input image to be transformed to the intermediate layer output, or (2) the ability to query the model to generate pairs of images and their corresponding intermediate output, such pairs can be used later to recover the model parameters. Our framework prevents this attack because the model parameters can only be decrypted inside the enclave that is rented by the user. The attacker (cloud provider) cannot see the model parameters that are executed inside the enclave. In addition, the user can add an authentication mechanism~\cite{yerba} to block any external adversaries from querying their enclave device which prevents the adversary from obtaining image and intermediate output pairs.



\section{Related Work}\label{sec:related}

{\bf Privacy Preserving Machine Learning:} Several approaches have been proposed to address the challenge of running machine learning (ML) workloads with reasonable accuracy while maintaining data confidentiality. Cryptographic-based~\cite{cryptonets} approaches are based on two main concepts: fully homomorphic encryption (FHE) and secure multiparty communication (SMC) which allows machine learning predictions to operate on encrypted without ever decrypting it and the output of the computation is the same as if it happened on unencrypted data. Although such systems achieve reasonable accuracy, their current performance makes them not practical for production environments. On-device inference~\cite{on-device1}\cite{on-device2} has also been proposed to perform ML in a privacy preserving manner through processing parts of the application in the client's device. However, on-device inference incurs high energy cost and significantly affects the lifetime of battery-operated devices. The approach that we adopt in this paper is to use Trusted Execution Environments (TEE) such as Intel SGX which have significant performance benefits over homomorphic encryption and do not incur the significant energy cost associated with on-device inference.

{\bf Machine Learning in SGX:} Several systems have proposed to run machine learning workloads in Intel SGX  to preserve data confidentiality~\cite{chiron}\cite{myelin}\cite{slalom}\cite{occulumency}\cite{oblivious}. However, none of them  have addressed the problem of partitioning computation across multiple enclaves or highlighted the tradeoff between using multiple enclaves vs. one enclave and another processor.
Occulmency~\cite{occulumency} and Myelin\cite{myelin} focus on accelerating the computation within one enclave. Occulmency proposes memory-efficient techniques for computing convolution and incrementally loading model parameters instead of loading the entire model at once. Serdab can benefit from such optimizations to accelerate the inference on the partial DNN deployed in each Enclave.
Chiron~\cite{chiron} focuses on training ML models while keeping the model parameters protected from the user, Serdab however focuses on inference rather than training and the model parameters are owned by the user and protected from the cloud provider. Yerbabuena~\cite{yerba} and Slalom~\cite{slalom} focus on offloading part of the computation to a colocated processor. Serdab goes beyond this work by offering partitioning across multiple enclaves and exploring the privacy tradeoffs that it entails.

{\bf Neural Network Partitioning:} Existing neural network inference systems, such as Clipper~\cite{clipper}, TensorFlow Serving~\cite{serving} focus mainly on ease of deployment, where the entire neural network is considered as a black box and deployed into one Docker container. However, they miss the opportunity of leveraging hierarchical clusters through deploying some layers of the neural network on the edge server and the rest of the layers in a remote cloud or another edge server. On the other hand, recent systems that have proposed placement of neural network operations across hierarchy of resources such as VideoEdge~\cite{videoedge}, and Neurosurgeon~\cite{neurosurgeon} \emph{did not consider partitioning a neural network for a stream of video frames which requires modeling pipeline parallelism}. Both systems base their partitioning/placement decisions on a single video frame rather than a stream of frames.

 


\section{Conclusion}\label{sec:conc}

In this paper, we present Serdab, a framework for analyzing video streams across multiple enclave devices to preserve data privacy. Serdab allows partitioning deep neural network layers across multiple devices.  
We leverage two insights to find the best placement of NN layers across devices. The first insight is that the intermediate output of shallow NN layers tends to be more similar to the original input than the output of deeper layers, hence the enclave can run only the shallow layers until the output is not correlated to the original input. We validate our findings by conducting a user study and we realize that users cannot figure out the objects in the image when the resolution of layer's output is less than 20x20 pixels. The second insight that we leverage in this work is pipeline parallelism which happens when both enclaves are concurrently processing different frames which improves the completion time compared to running the majority of the workload in one enclave. Our results show that for a stream of 10,800 frames, our partitioning strategy achieves up to 4.7x speedup compared to executing the entire neural network in one enclave.


\section{Acknowledgements}\label{sec:ack}
 This research is supported and funded by The Aerospace Corporation’s University Partnership Program. We thank Mikhail Tadjikov, John Maguire, and Ingrid Guch who are members of The Aerospace corporation for their valuable feedback on this work.


\end{document}